\documentclass[twocolumn]{emulateapj}
\submitted{Science with Parkes @ 50 Years Young, 31 Oct. -- 4 Nov., 2011}

\usepackage{graphicx}

\shorttitle{Fifty Years in Fifteen Minutes}
\shortauthors{Edwards}

\begin{document}

\title{Fifty Years in Fifteen Minutes: \\
The Impact of the Parkes Observatory}

\author{Philip Edwards}
\affil{CSIRO Astronomy \& Space Science, PO Box 76, Epping 1710 NSW, Australia}
\email{Philip.Edwards@csiro.au}

\begin{abstract}
The scientific output of Parkes over its fifty year history is briefly 
reviewed on a year-by-year basis, and placed in context with other 
national and international events of the time.
\end{abstract}



\section{Introduction}

The Sydney Morning Herald published a cartoon by George Molnar after the
opening of the Parkes telescope in 1961, in which a character on horseback,
looking at the telescope, explains to a mate ``It's the telescope of the 
future. It can look back millions of years.''
(The cartoon is reproduced in Robertson 1993.)

In the weeks before the Parkes 50th symposium I happened to read 
(in the Sydney Morning Herald!) Danish philosopher S{\o}ren Kierkegaard's
statement 
``Life can only be understood backwards, but it must be lived forwards.’’ 

This paper, based on a presentation at the Parkes 50th Symposium, attempts
to combine these viewpoints to look back over the preceding 5 decades 
to determine how \lq the telescope of the future' has contributed to 
the development of astronomy by selecting a small number of highlights or 
incidents from each year, and placing them in the context of other
national and international events of note from the time.

In most cases, the paper selected for each year is the one which
has had the greatest impact, as assessed by the number of
subsequent citations amassed.
This work has made extensive use of the SAO/NASA Astrophysics Data
System (ADS), and it is worth repeating their caveat that 
``The Citation database in the ADS is NOT complete. 
Please keep this in mind when using the ADS Citation Lists.''
It should be noted that ADS citation counts are
less accurate in the first decades of the Observatory's existence,
and also that searching for ``Parkes'' in the abstract of papers will 
inevitably miss many papers which appeared in journals
such as Nature and Science... but will pick up many papers not 
reporting the results of observations with the Parkes telescope(s).
As a result, this paper does not purport to be a complete listing 
of the highest impact papers, but does endeavour to illustrate both the 
nature and breadth of high impact research conducted at the
Observatory.

\section{Observations}

\subsection{1961}

October 31st saw the official opening of the
Parkes 210-foot radio-telescope, and commissioning
work undertaken.
John Bolton returned from Owens Valley to become Officer-in-Charge (OiC) of the 
Australian National Radio Astronomy Observatory (ANRAO).
The world's population passed 3 billion, 
Yuri Gagarin orbited the Earth, the (first version of the) 
Berlin Wall was constructed,
and Joseph Heller's {\em Catch 22} was published.

\subsection{1962}

Telescope commissioning ended in 1962 and early observations yielded
a number of fundamental results. 
In ``Polarization of 20-cm Wavelength Radiation From Radio Sources''
Gardner \& Whiteoak (1962) noted that their observations
of linear polarization ``...considerably strengthens the hypothesis
that the synchrotron mechanism is responsible for the radiation
from the nonthermal sources.''
Observations of this linear polarization as a function of frequency
quickly resulted in the detection of Faraday rotation:
``Faraday Rotation Effects associated with the Radio Source Centaurus A''
(Cooper \& Price 1962) and
``Polarization in the Central Component of Centaurus A''
(Bracewell et al. 1962).
Elsewhere, John Glenn orbited the Earth, Marilyn Monroe died,
the Cuban missile crisis was played out,
and Rod Laver won all four tennis Grand Slam tournaments in the same calendar year.

\subsection{1963}

A series of occultations of 3C273 by the moon enabled
the location of this bright radio source to be located
with sufficient accuracy for its optical counterpart to be identified:
as a result 3C273 and 3C48 became the first recognised quasars: 
``Investigation of
the Radio Source 3C 273 By The Method of Lunar Occultations'' (Hazard,
Mackey \& Shimmins 1963).  Gardner \& Whiteoak (1963) continued their
polarisation studies; ``Polarization of Radio Sources and Faraday
Rotation Effects in the Galaxy'', inferring a galactic magnetic field
from the measurements of Faraday Rotation as a function of galactic
coordinates.  In ``A Radio Source with a Very Unusual Spectrum,''
Bolton, Gardner \& Mackey (1963) presented the first study of the
source which would become the ATCA's primary flux density calibrator
at cm-wavelengths, 1934$-$638.

The year started with the Bogle \& Chandler mystery: the discovery on
New Year's Day of the bodies of (CSIRO scientist) Gilbert Bogle \&
Margaret Chandler, and in November US President John F.\ Kennedy was
assassinated.


\ 

\subsection{1964}

The first zone of 408\,MHz survey was published (Bolton,
Gardner \& Mackey 1964), and
following the discovery of the OH main lines at 1665 and 1667 MHz,
Gardner, Robinson, Bolton \& van Damme (1964) reported 
``Detection of the Interstellar OH Lines at 1612 and 1720 Mc/sec''.
The Beatles toured Australia, and the summer Olympics were held in Tokyo,
with Dawn Fraser and Betty Cuthbert among Australian gold medal winners.

\subsection{1965}

In 1965 the Kennedy 60-foot (18-m) antenna became operational.
The antenna was built by the company founded by Donald Snow Kennedy in Cohasset, 
Massachusetts\,\footnote{See http://www.wickedlocal.com/cohasset/news/x1001333925/The-business-of-antennas 
for more details.}
The company operated from 1947 to 1963, 
with 
an advertisement in the September 1956 {\em Scientific American} 
promising
``Down-To-Earth Solutions to Out-Of-This-World Problems'',
and another 60-foot telescope becoming 
the George R.\ Agassiz Radio Telescope of Harvard Observatory
(Bok 1956).


In a one-page paper, ``The supernova of A.D. 1006'', Gardner and Milne (1965)
solved a 959 year old mystery by identifying the remnant of SN1006 with
the polarized, extended radio source 1459$-$41.

The Tidbinbilla Deep Space Tracking Station and was officially opened 
by Prime Minister Sir Robert Menzies; who in the same year
committed Australian troops to Vietnam and reintroduced conscription.

\subsection{1966}

The ``21 cm hydrogen-line survey of the Large Magellanic Cloud. II. 
Distribution and motions of neutral hydrogen'' of McGee \& Milton (1966)
was carried out over several years with a 48 channel line receiver
with a spectral resolution of 7\,km\,s$^{-1}$.

Kellermann's (1966) discovery of radio emission from Uranus ---
which is the ATCA's primary flux density calibrator in the mm bands --- 
meant that with cm and mm calibrators identified it would only take 
another 25 years for the telescope that would use them to come along!

The Astronomical Society of Australia (ASA) was founded, with Harley
Wood as its first President.  The business of finding optical
counterparts to radio sources was booming: PKS 0106+01, identified by
Bolton, Clarke, Savage \& V{\'e}ron (1965) 
became the most distant object then known, with
a redshift of $z$=2.1 being determined by Burbidge (1966).

Decimal currency was introduced in Australia, replacing pounds and pence.
St Kilda won the Victorian Football League (VFL) Grand Final and 
England won the soccer/football World Cup: neither feat has been repeated!

\begin{figure*}
\centering
\includegraphics[width=100mm]{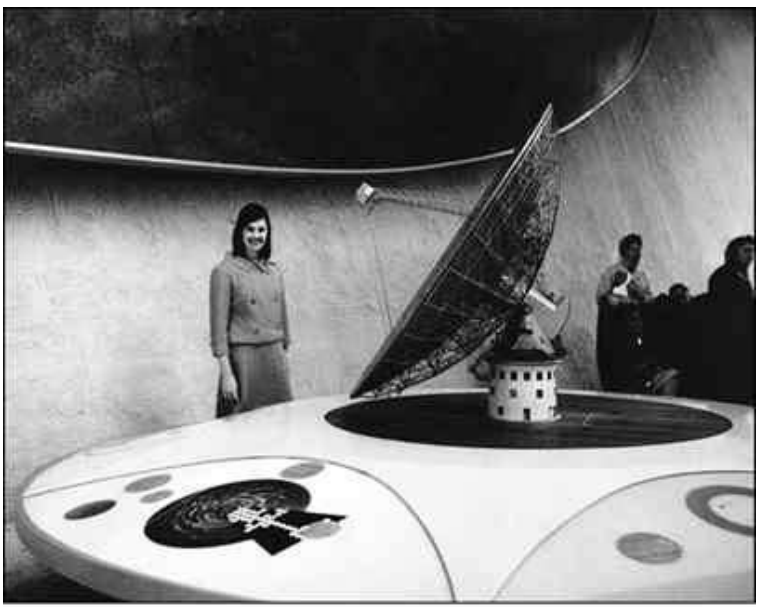}
\caption{Model of the Parkes radio telescope in the Australian Pavilion at Expo ’67, held in Montreal. The model is now part of the Parkes Visitor Centre.
(Courtesy of National Archives of Australia [AA1982/206, 44])} \label{f1}
\end{figure*}

\subsection{1967}

The 48-channel spectrometer was also used by Kerr \& Vallak (1967)
in presenting their
``Hydrogen-Line Survey of the Milky Way I. The Galactic Center''.
The farthest known object became another Parkes source, 
PKS0237$-$23 at $z$=2.22 (Arp, Bolton \& Kinman 1967).
The Byrds released their LP record {\em Younger than Yesterday},
which is notable for the song ``CTA 102'', written following
press reports of speculation that this radio source contained transmissions from an
extra-terrestrial civilisation (see Kellermann 2002 for more details).
The Molonglo Cross radiotelescope began full operation at 408\,MHz.

The World Fair was held in Montreal: Barnes \& Jackson (2008) note
that ``Expo '67 marks Australia’s return to international exhibitions
after nearly thirty years. Planned during the period of economic
disengagement from Britain, the pavilion reveals how the register for
Australia's self-representation had unfolded since 1939. Australia now
emphasised its scientific and technical proficiency with large-scale
models of the Snowy Mountains Hydroelectric Scheme and the Parkes
radio telescope, as well as evidence of manufacturing capacity through
examples of modernist furniture and product design'' (see Figure~1).

In December, Prime Minister Harold Holt disappeared while swimming,
with John McEwen subsequently becoming Prime Minister.

\subsection{1968}

The 408\,MHz survey completed its northernmost zone,
+20$^\circ<$dec$<$+27$^\circ$ (Shimmins \& Day 1968) with follow-up
optical identifications in close pursuit (Bolton, Shimmins \&
Merkelijn 1968).  The discovery of pulsars was announced and Parkes'
first contribution was to correct the 5th significant figure of the
period of the first pulsar: ``Measurements on the Period of the
Pulsating Radio Source at 1919+21'' (Radhakrishnan, Komesaroff, Cooke
1968).  

The Prague Spring saw political liberalisation begin in
Czechoslovakia in January, and end in August with the invasion by the
Soviet Union.  Martin Luther King was assassinated in April and Robert
Kennedy in June.  The summer Olympics were held in Mexico City, with
Michael Wenden winning the 100\,m and 200\,m freestyle events.

\subsection{1969}

Radhakrishnan \& Cooke (1969) described 
``Magnetic Poles and the Polarization Structure of Pulsar Radiation'',
and Parkes observations of the first Vela glitch were reported
(though not in those words) in
``Detection of a Change of State in the Pulsar PSR 0833$-$45''
(Radhakrishnan \& Manchester 1969).

The first VLBI observations with Parkes were undertaken in 1969,
with fringes on the baseline to Owens Valley being found from
observations in April (Kellermann et al.\ 1971).
One of the many technical challenges was described:
``Time in Australia was synchronized with time in Owens Valley via
the NASA tracking stations at Tidbinbilla, Australia, and Goldstone,
California. The tracking stations themselves were synchronized
to an accuracy of a few microseconds by
the transmission of a radar signal from Goldstone to Tidbinbilla 
via the moon.''

Apollo 11 was launched on July 16 (UT), landed on the moon on July 20, 
with the ``small step and giant leap'' relayed via Parkes on July 21
(Sarkissian 2001). Apollo 12, which followed in November the same year,
was supported by the team pictured in Figure~2. 


\subsection{1970}

The discovery of recombination lines is described by 
Robinson (1994): Parkes may have missed the opportunity
to have first observed these but was quick to follow-up,
with 
``A Survey of H 109$\alpha$ Recombination Line Emission in 
Galactic H\,{\sc ii} Regions of the Southern Sky'' 
(Wilson, Mezger, Gardner \& Milne 1982) being 
made with the NRAO 6\,cm cooled receiver.
The Metric Conversion act is passed, instantly 
converting the 210-foot telescope
into a 64\,m telescope!
The first stage of resurfacing the dish with perforated aluminium panels 
started.
Apollo 13 was launched on April 11 (UT), limping back to Earth 6 days later,
again with significant, and hastily arranged, Parkes contributions 
(e.g., Bolton 1994; Sherwen 2010).

\subsection{1971}

The Parkes 2700 MHz Survey
was in full swing, with 
``Catalogues for the $\pm 4 ^\circ$ declination zone and for the selected regions'' published by Wall, Shimmins \& Merkelijn (1971).
Bolton stepped down as OiC in 1971, with John Shimmins taking on the role, 
however Bolton stayed on as ``Astronomer at Large'' until 1981.
%
%
%

The year also saw first McDonalds opened in Australia, the
South Sydney Rabbitohs win the rugby league grand final, and
Evonne Goolagong win Wimbledon.

\subsection{1972}

The first five papers (and 166 pages) of the Astrophysical Journal
Supplement volume 24 reported results from
the Parkes Hydrogen Line Interferometer, in which
the ``signal from a remotely controlled movable 18-m paraboloid is cross-correlated with the signal from the stationary 64-m reflector.''
The papers were all titled
``The Parkes Survey of 21-cm Absorption in Discrete-Source Spectra''
with paper I describing the Parkes Hydrogen-Line Interferometer
(Radhakrishnan, Brooks, Goss, Murray, \& Schwarz 1972);
II. Galactic 21-cm Observations in the Direction of 35 Extragalactic Sources
(Radhakrishnan, Murray, Lockhart, \& Whittle 1972);
III. 21-Centimeter Absorption Measurements on 41 Galactic Sources North of Declination $-$48$^\circ$ (Radhakrishnan, Goss, Murray \& Brooks 1968);
IV. 21-Centimeter Absorption Measurements on Low-Latitude Sources South of Declination $-$46$^\circ$ (Goss, Radhakrishnan, Brooks \& Murray  1968)
; and 
V. Note on the Statistics of Absorbing H\,{\sc i} Concentrations in the Galactic Disk (Radhakrishnan \& Goss 1968).

The second stage of resurfacing the dish with panels 
to a diameter of 37\,m was carried out.
Gough Whitlam became Prime Minister; the summer Olympics were overshadowed by 
the ``Munich massacre'' of 11 Israeli athletes, coaches and officials.

\subsection{1973}

The growing number of discoveries of organic molecules in space
led to a collaboration between Radiophysics astronomers and
Monash University chemists, resulting in 
the discoveries of interstellar methanimine, CH2NH, at 5290\,MHz
(Godfrey, Brown, Robinson, \& Sinclair 1973),
and thioformaldehyde, CH2S, at 3139\,MHz
(Sinclair, Fourikis, Ribes, Robinson, Brown \& Godfrey 1973).

The IAU General Assembly adopted the jansky as the unit of measurement
for spectral flux density.
The Australian \$50 note was first issued, with Ian Clunies-Ross 
(first chairman of the re-named CSIRO) pictured on one side, together 
with an image of the Parkes radio telescope, pulses from CP1919, and 
a radio image of the LMC.
The Tidbinbilla 64m telescope was completed, and 
in October the Queen officially opened the Sydney Opera House.

\begin{figure*}
\centering
\includegraphics[width=100mm]{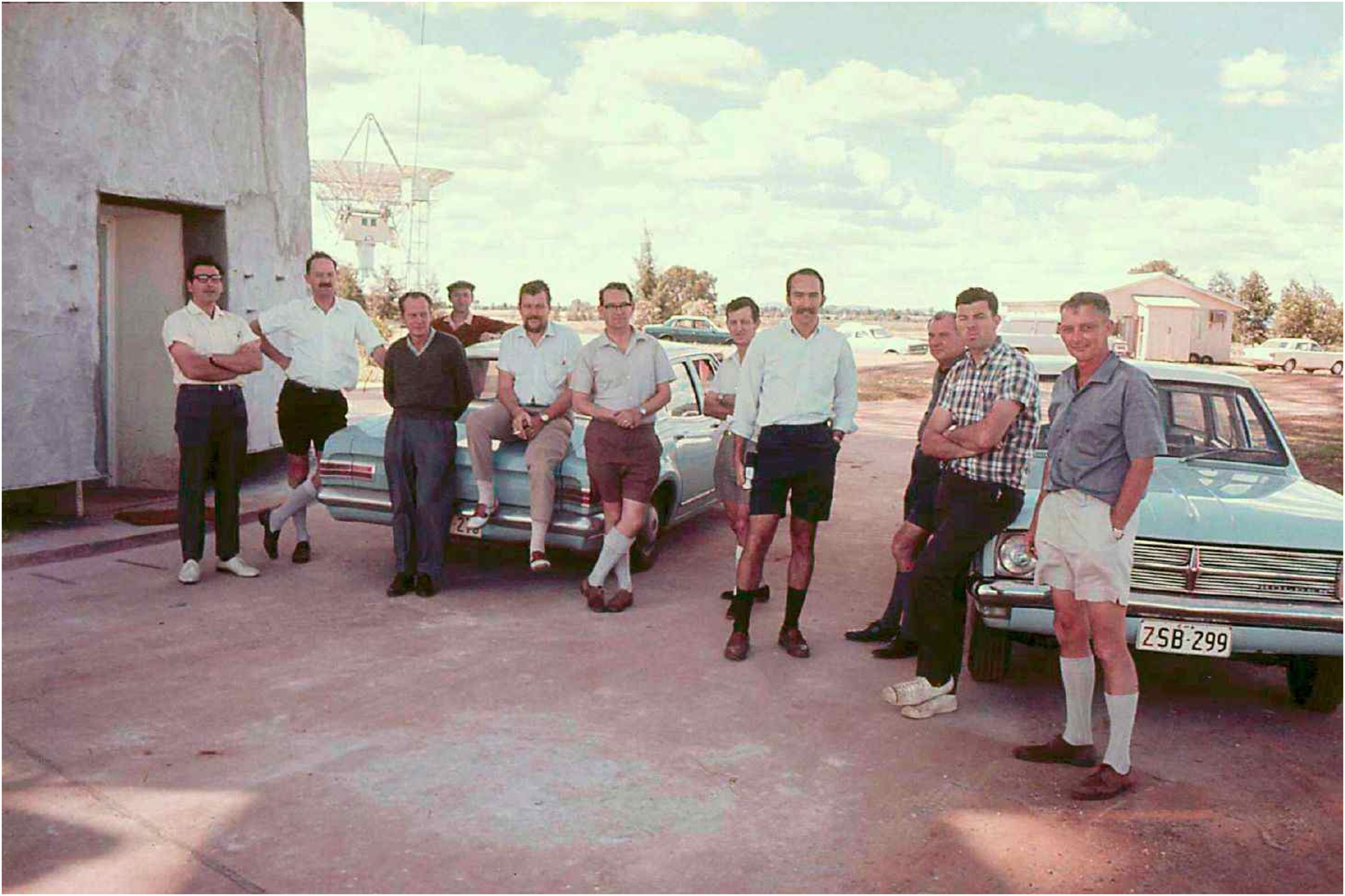}
\includegraphics[width=30mm]{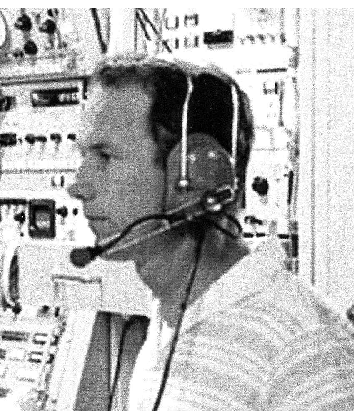}
\caption{Support crew for the Apollo 12 tracking at Parkes.
The Observatory staff members pictured are
Dennis Gill (third from left), John Shimmins (fourth from left),
Dave Cooke (sixth from left) and John Bolton (far right).
The telescope tower had a concrete jacket added between the
Apollo 11 and 12 missions to strengthen the structure.
(Picture credit: Bruce Window, courtesy of honeysucklecreek.net)
The insert at right is of Bruce Window, who led the team
from Tidbinbilla for the Apollo 12 support (Picture credit: CSIRO).
} \label{f2}
\end{figure*}

\subsection{1974}

A southern sky survey of neutral hydrogen with the Parkes 18\,m
revealed a long filament of H\,{\sc i} extending from the Magellanic Clouds
to the south galactic pole, which was dubbed
``The Magellanic stream'' (Mathewson, Cleary \& Murray 1974). 
Extragalactic formaldehyde was discovered in absorption by Gardner \& Whiteoak (1974),
using the 6cm cryogenic parametric amplifier  
and a 512 channel autocorrelator.
Prince Charles inaugurated the Anglo-Australian Telescope.
The world population passed 4 billion, Abba won the Eurovision Song Contest,
and Cyclone Tracy devastated Darwin on December 24th.

\subsection{1975}

Caswell, Murray, Roger, Cole \& Cooke (1975)
used the Parkes hydrogen line interferometer to make
``Neutral Hydrogen absorption measurements yielding kinematic 
distances for 42 continuum sources in the galactic plane'';
about half the sources were supernova remnants and most of the remainder were H\,{\sc ii} regions. 
The first VLA antenna was put in place.
Radio Astronomy was commemorated on a 24c Australian Stamp:
``Radio Astronomy explores and maps the heavens by examining the radio waves emitted by stars, 
galaxies and gas clouds. Discovery \& development of unique forms of radio telescopes have 
been contributed by Australia.''
Malcolm Fraser became Prime Minister after the ``constitutional crisis''.

\subsection{1976}

Clark \& Caswell (1976) presented
``A study of galactic supernova remnants, based on Molonglo-Parkes observational data'', inferring
a larger characteristic interval between galactic supernovae, $\sim$ 150 years, than had previously been assumed. 
The first commercial flight of a Concorde was made.
The summer Olympics were held in Montreal, with Australia winning just 
a single silver medal and four bronze medals.

\subsection{1977}

Allen et al.\ (1977) published
``Optical, infrared, and radio studies of AFCRL sources.'' 
The acronym AFCRL needed no expansion at the time (and none is given 
in the paper!) but it is probably now necessary to note that 
AFCRL sources are sources detected in the infra-red surveys
at 4, 11 and 20 $\mu$m 
conducted by the Air Force Cambridge Research Laboratories.
The Parkes observations were made at the 1612\,MHz OH line.
The Apple company was incorporated.
In January, 83 people died in the Granville train disaster when a 
Sydney-bound commuter train derailed and collided with the supports
of a road bridge, which collapsed onto the train.
In August, Elvis Presley died (or did he? 
Numerous sightings are reported in Parkes every January, adding
credence to rumours that in fact he lives on...).

\subsection{1978}

An area of approximately 600 square degrees was surveyed by 
Haynes, Caswell \& Simons (1978) for
``A southern hemisphere survey of the galactic plane at 5 GHz.''
The second Molonglo pulsar survey was published.
The Sydney Hilton bombing and Jonestown massacre made headlines.

\subsection{1979}

Caswell \& Lerche (1979) described
``Galactic supernova remnants --- Dependence of radio brightness on galactic height and its implications'' 
and resolved an anomaly concerning the faint (but young) remnant of SN1006.
The 14th and final part of 2.7\,GHz survey was published (Bolton et al.\ 1979).
Margaret Thatcher was elected, and the Iran hostage crisis began.

\subsection{1980}

Batchelor et al.\ (1980) reported on  
``Galactic plane H$_2$O masers --- A southern survey,''
observations for which were made with Parkes 17m and 37m telescope 
(i.e., illuminating the inner 17m of the 64m for the first epoch, in 1975, and the inner panelled 37m for the later epochs).
The Two Element Synthesis Telescope (TEST) was instituted to test and gain experience
as design work commenced for the ATCA. 
Parkes VLBI efforts were renewed with participation in the 
2.3 GHz all sky VLBI survey of Preston et al.\ (1985). 

The VLA was inaugurated, the Pac-man video game was released, and Azaria Chamberlain disappeared at Ayers Rock.
The summer Olympics were held in Moscow, with a boycott being led by the USA
in response to the Soviet invasion of Afghanistan.
Australian athletes were encouraged not to attend: a reduced team
did participate but marched behind the Olympic flag rather than 
the Australian flag at the Opening Ceremony.


\subsection{1981}

The  
``Simultaneous X-ray, ultraviolet, optical, and radio observations 
of the flare star Proxima Centauri'' (Haisch et al.\ 1981)
was an early an example of a coordinated
multi-wavelength campaign on an astronomical object. 
With the retirement of John Shimmins, Jon Ables became OiC.
The wedding of Prince Charles and Diana Spencer was seen by a global television audience of over 750 million.

\subsection{1982}

Haslam, Salter, Stoffel, \& Wilson (1982)
published ``A 408\,MHz all-sky continuum survey. II --- The atlas of contour maps'' based on data from four different surveys using the Jodrell Bank MkI, Bonn 100\,m, Parkes 64\,m and Jodrell Bank MkIA telescopes.
Parkes wrested back the record for the most distant known object 
with the discovery that PKS 2000$-$330 lay at a 
redshift of $z$=3.78 (Peterson, Savage, Jauncey \& Wright 1982).
Perforated aluminium panels were extended to 44\,m diameter.
The Falklands War started in April and ended in June.
The Commodore 64 was released, and the year also saw the
first use of emoticons on the internet :-)

\subsection{1983}

Shaver, McGee, Newton, Danks \& Pottasch (1983) used radio and optical spectroscopy
to study chemical abundances in a sample of 67 H\,{\sc ii} regions covering a wide range of
galactocentric radius and derive ``The galactic abundance gradient.''
The Parkes observations were made with the 6cm cryogenic receiver and the 512-channel correlator.
The first extra-galactic pulsar was discovered, using the 64m, in the LMC
(McCulloch, Hamilton, Ables \& Hunt 1983).
Bob Hawke become Prime Minister, and Australia won the
America's Cup with the Ben Lexcen-designed winged keel.

\subsection{1984}

Rohlfs, Kreitschmann, Feitzinger, \& Siegman (1984) published
``A neutral hydrogen line survey of the Large Magellanic Cloud'' 
made with the 64m telescope and a channel width of 0.8\,km\,s$^{-1}$.
Medicare came into effect, and Advance Australia Fair became the national anthem.
The \$1 coin was introduced, replacing the \$1 note.
The Los Angeles Olympics were held, with the Soviet Union leading
a boycott.


\subsection{1985}

Manchester, D'Amico \& Tuohy (1985) conducted
``A search for short-period pulsars'' 
at 1.4\,GHz (to minimize interstellar scattering) 
with a sampling interval of 2\,ms, 
in a targeted survey of supernova remnants and 
gamma-ray sources, detecting four new pulsars.
Windows 1.0 was released.

\subsection{1986}

The
``Fully sampled neutral hydrogen survey of the southern Milky Way'' of Kerr,
Bowers, Jackson \& Kerr (1986) 
was carried out with the Parkes 18m at a velocity resolution of 2.1\,km\,s$^{-1}$.
The Parkes 64\,m supported the Voyager fly-by of Uranus for NASA, and Giotto at 
Halley's comet for ESA.  
An Australian 33c stamp pictured the Parkes radio telescope in its
support role of missions to Halley's comet.  The perihelion pass of
Halley's comet generated a lot of public interest.  Ecos vol.\ 47
indicates how these were managed in a pre-internet world: ``For a
weekly update on the comet and associated phenomena, dial Sydney or
Brisbane 11622 or Melbourne 11613. The weekly scripts for this joint
Telecom-CSIRO community service are prepared by Dr Ray Norris of the
CSIRO Division of Radiophysics.''

The Mt Pleasant 26\,telescope (formerly used as a tracking station antenna at Orroral Valley)
was formally opened during the annual ASA meeting, held in Hobart in May.
The Space Shuttle Challenger broke up 73 seconds after launch, killing all seven crew, 
and the Chernobyl nuclear reactor disaster occurred in April.

\subsection{1987}

Caswell \& Haynes (1987) tabulated 316 
``Southern H\,{\sc ii} regions --- an extensive study of radio 
recombination line emission'' 
observed in the 5\,GHz band.
SN1987A was discovered in February
and the Parkes--Tidbinbilla Interferometer played a role in follow-up 
observations in the days after the supernova
(e.g., Turtle et al.\ 1987).
The Tidbinbilla 64\,m was expanded to a diameter of 70\,m.
The world population exceeded 5 billion.

\subsection{1988}

Mathewson, Ford \& Visvanathan (1988) combined optical, infrared and
Parkes H\,{\sc i} observations to study
``The structure of the SMC. II'', finding evidence to support the
hypothesis that the SMC was disrupted by a collision with the LMC some
2$\times10^8$ years ago.
 
The Australian Bicentennial year was celebrated (in part) with the
official opening of the Australia Telescope Compact Array 
in September (again, held in conjunction with the annual meeting of the ASA).
The birth of one telescope was however coincided with the 
passing of another, with the collapse of the NRAO 300-foot dish in November.
A Memorandum of Understanding was signed between the CSIRO 
Office of Space Science and Applications (COSSA) and the USSR Academy of Sciences
for participation in 
the RadioAstron space VLBI mission.
CSIRO was also involved in the research behind Australia's first 
polymer banknotes, which were released this year.
The Olympic games were held in Seoul.

\subsection{1989}

Aaronson et al.\ (1989) used the new (in 1986) multi-band receiver
developed as a prototype for the ATCA to detect galaxies in
six clusters of galaxies, displaying
``Large peculiar velocities in the Hydra-Centaurus supercluster.'' 
Internet connections were established to most Australian Universities.
At Parkes, the old VAX 11/750 was replaced by a MicroVax 3400
costing \$172,000.
Dave Cooke became OiC, the
Berlin Wall came down, and Australian pilots went on strike, 
severely impacting domestic travel.

\ 

\ 

\subsection{1990}

The PKSCAT90 version of the Parkes Catalog was released (Wright \&
Otrupcek 1990), containing radio and optical data for the 8264 radio
sources in the Parkes 2700 MHz Survey, covering all the sky south of
a declination of +27$^\circ$ but largely excluding the Galactic Plane and
the Magellanic Cloud regions.
The 4850 MHz receiver that had been on the NRAO 300-foot telescope 
at the time of its collapse was brought to Parkes for the Parkes-MIT-NRAO (PMN)
surveys, carried out in June and November 1990, covering the sky between
$-87^\circ<$ Dec. $<+10^\circ$.
The Australia Telescope National Facility was established,
and the Hubble Space Telescope was launched.
Nelson Mandela was released after 27 years incarceration.

\subsection{1991}

te Lintel Hekkert et al.\ (1991) conducted a ``1612 MHz OH survey of
IRAS point sources. I --- Observations made at Dwingeloo, Effelsberg
and Parkes''.  A total of 2703 IRAS sources were observed, with 738
OH/IR stars being detected, 597 of which were new discoveries.  (The
first) Ten millisecond pulsars were discovered in the globular cluster
47 Tucanae (Manchester et al.\ 1991).  

Ron Ekers planted an apple tree
near the (then) entrance to the visitors centre.  The tree was a
direct descendent of the apple tree which is reputed to have
stimulated Newton's development of a theory of gravitation.  The tree
has struggled with drought (and being run over!) but is now
complemented by additional trees in the garden outside the new VC
entrance, with signage telling the story behind the tree's arrival at
Parkes.

The year also saw first light with Mopra, 
the public debut of the world wide web, the launch of the Compton Gamma-Ray Observatory,
and Paul Keating become Prime Minister.

\subsection{1992}

Johnston et al.\ (1992) reported the discovery of 
``PSR 1259$-$63 --- A binary radio pulsar with a Be star companion.'' 
The 47\,ms pulsar is in a highly eccentric orbit around its massive companion,
with the pulsar eclipsed by the companion star's stellar wind near periastron.
The high-frequency (1500\,MHz --- high frequency for pulsar astronomers!)
survey of 800 square degrees of the southern Galactic plane that yielded
the discovery of PSR 1259$-$63 was also published (Johnston et al.\ 1992b).
One-person-in-the-tower operation of the telescope commenced.
``Beyond Southern Skies: Radio Astronomy and the Parkes Telescope'' 
(Robertson 1992) was published.
The astro-ph server started, and the NASA/ADS website was launched.
The summer Olympics were held in Barcelona, with Kieren Perkins winning the
1,500\,m freestyle, the Oarsome Foursome winning the men's coxless fours,
and Australia earning the gold medal in the Equestrian three-day team event.

\subsection{1993}

The first PMN paper, 
``The Parkes-MIT-NRAO (PMN) surveys. I --- The 4850 MHz surveys and data reduction'' 
was published (Griffith \& Wright 1993).
One of the limitations of the survey was recognised after the observations as being
due to ``Complex, off-axis sidelobes of a radio telescope caused by feed-support legs''
(Hunt \& Wright 1992) which allowed enough radiation from the sun to enter the
feed to compromise a small part of the surveyed area.
The European Union was established, and Bill Clinton became US President.

\subsection{1994}

The second PMN paper:
``... Source catalog for the southern survey  $-87.5^\circ<\delta<-37^\circ$'' 
was published (Wright, Griffith, Burke \& Ekers 1994).
Marcus Price became OiC, however CSIRO budget cuts resulted in 
six staff being made redundant at Parkes during the year.
``Parkes, Thirty Years of Radio Astronomy'' (Goddard \& Milne 1994)
was published.
The premature deaths of 
Kurt Cobain and Ayrton Senna were mourned by the music and sporting worlds, respectively.

\subsection{1995}

``Relativistic motion in a nearby bright X-ray source'' was
reported by Tingay et al.\ (1995), based on 
Target of Opportunity VLBI observations 
of GRO J1655-40 over four days in August.
The SHEVE (Southern Hemisphere VLBI Experiment) array 
included Parkes, Tidbinbilla, Hobart, Mopra and ATCA.

Project Phoenix used Parkes (as the primary station) and Mopra
(for rapid independent follow-up of candidates) from February to June
(Tarter 1997).
Observing was shut-down for two months for the new focus cabin installation. 
The DVD format was announced and eBay was founded.
The former has been deployed in archiving telescope data,
and the latter has proved useful in sourcing otherwise hard-to-find components
for more than one piece of astronomical equipment!

\subsection{1996}

This year saw publication of ``The Parkes 21 cm multibeam receiver''
(Staveley-Smith et al.\ 1996).
The paper does not present any results from Parkes publications 
(so in the strictest sense should not qualify for inclusion)
but is notable for the large number of citations, and for
describing what went on to become what is probably the Observatory's
most productive receiver.
The paper concludes:
``Documentation of the Parkes multibeam receiver, including more details on
scientific goals, observing and data-reduction techniques, can be found on the
World Wide Web. The address (as of July 1996) is
http://www.atnf.csiro.au/Research/multibeam/ multibeam.html.''
The foresight of acknowledging that URLs may not be permanent was
well-founded, as that address now yields a ``404'', however 
http://www.atnf.csiro.au/research/multibeam/
multibeam.html
does (at the time of writing!) still exist.
The dish spent much of the year tracking  Galileo,
which arrived at Jupiter in December 1995, for NASA.

Henry Parkes' image appears on the Australian
one-dollar coin of 1996, commemorating the 100th anniversary of his death.  
(It is worth noting that the 
original settlement in 1853 was named Currajong,
which later became Bushman's Lead, or simply Bushman's. 
It was not until 1873 that the town was renamed Parkes, after the then Premier of NSW.)

John Howard became Prime Minister.
The summer Olympics were held in Atlanta, with 
back-to-back gold medals for Kieren Perkins, the Oarsome Foursome,
and the Equestrian team!


\subsection{1997}

C{\^o}t{\'e}, Freeman, Carignan \& Quinn (1997)
scanned SRC J films to find dwarf irregular galaxy candidates in the 
Sculptor and Centaurus groups of galaxies, and obtained redshifts 
with Parkes H\,{\sc i} and optical H$\alpha$ observations to report the 
``Discovery of Numerous Dwarf Galaxies in the Two Nearest Groups of Galaxies''. 
Galileo tracking continued, and the MB20 receiver was 
installed for the first time.
The HALCA satellite of the VLBI Space Observatory Programme (VSOP) was launched,
the first Harry Potter novel published, Princess Diana died, and 
the Adelaide Crows won the AFL Grand Final.

\subsection{1998}

The first results from 20\,cm multibeam receiver observations included 
``Tidal disruption of the Magellanic Clouds by the Milky Way'' 
by Putman et al.\ (1998), which revealed a stream of
atomic hydrogen leading the motion of the clouds (i.e., on the opposite
side of the Magellanic stream).
John Reynolds became OiC, and the GST (Goods \& Services Tax) was introduced.

\subsection{1999}

Stanimirovic, Staveley-Smith, Dickey, Sault, \& Snowden (1999) combined
single-beam Parkes 21\,cm data from 1996 with an ATCA mosaic to study
``The large-scale HI structure of the SMC.'' 
The world population topped 6 billion, the Euro currency was established,
the Mars Climate Orbiter was lost, and it was feared that
``Y2K'' would wreak havoc on computers.


\begin{figure*}
\centering
\includegraphics[width=100mm]{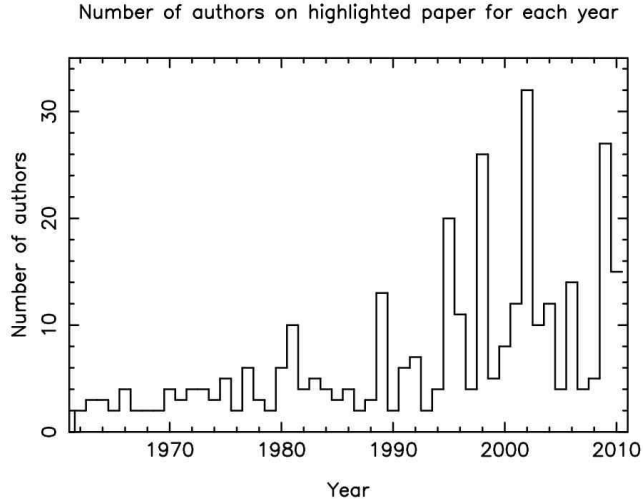}
\caption{The number of authors on the highlighted paper for each year,
illustrating the trend to larger collaborations with time.} \label{f3}
\end{figure*}

\subsection{2000}

The ``Discovery of Two High Magnetic Field Radio Pulsars'' 
with the multibeam receiver was reported by Camilo et al.\ (2000).
The Sydney Olympics were held, with the the olympic torch 
receiving a ride on the dish as the torch relay made its way past Parkes.
The movie ``The Dish'' was the top-grossing movie 
in Australia for the year, and the expanded Visitor Centre opened
just in time to welcome the increased numbers inspired to visit by the movie!

\subsection{2001}
This year marked the Centenary of Federation, with 
Henry Parkes' role commemorated by his picture appearing on a special \$5 note.
The first large ``Parkes Multi-beam Pulsar Survey'' (Manchester et al.\ 2001) 
and ``HIPASS'' (Barnes et al.\ 2001) papers appeared.
Four passenger jets were hijacked in the ``9/11'' terrorist attacks.

\subsection{2002}

A reprocessing of HIPASS data by Putman et al.\ (2002) yielded 1956 
High Velocity Clouds and
allowed a study of the
``... Properties of the Compact and Extended Populations.'' 
The ``10/12'' Bali bombings killed 202 people, and injured a further 240.

\subsection{2003}

Greenhill et al.\ (2003) made 22\,GHz VLBI observations with
Parkes, Hobart, Mopra and Tidbinbilla in 1997 and 1998 (before the ATCA had 
22\,GHz receivers) to observe maser emission tracing out
``A Warped Accretion Disk and Wide-Angle Outflow in the Inner Parsec of the Circinus Galaxy.'' 
The Parkes telescope was fitted with 180 new aluminium panels out to a diameter of 55\,m. 
The new 10/50-cm \& Mars (8.4\,GHz) receivers were delivered to the Observatory,
and the new wideband (up to 1\,GHz bandwidth) correlator put into service.
The 20\,cm multibeam was brought down from the focus cabin for the first time for 
repairs and refurbishment.
Mars tracking for NASA began in September. 
The IAU General Assembly in Sydney in July.
``First bite'' occurred at the Dish Caf{\'e}.
The Concorde made its Last flight, the
Space Shuttle Columbia disaster occurred, and
Canberra bushfires resulted in the destruction of much of the 
Mount Stromlo Observatory.

\subsection{2004}
Lyne et al.\ (2004) reported the discovery of 
``A Double-Pulsar System: A Rare Laboratory for Relativistic Gravity and Plasma Physics.''
The Boston Red Sox won the World Series, breaking an 86 year drought.
The Sumatra earthquake and tsunami on December 26th killed over 230,000 people.

\subsection{2005}

Zwaan,  Meyer, Staveley-Smith, \& Webster (2005)
used the catalogue of 4315 extragalactic HI 21-cm emission-line detections from HIPASS
to study 
``... $\Omega_{\rm HI}$ and environmental effects on the H\,{\sc i} mass function of galaxies,''
finding tentative evidence for a steeper mass function toward higher density regions. 
The ATNF Pulsar catalogue was published (Manchester et al.\ 2005).
Huygens arrived at Titan, with Parkes and Mopra joining an international network of radio telescopes
listening in (at 2040\,MHz) on the descent of the probe to the Saturnian satellite's surface.
Galactic All-Sky Survey (GASS) observing started in January.
The Sydney Swans won the AFL Grand Final, and Liverpool 
won the Champions League Final.

\subsection{2006}
The discovery of ``RRATs'' was reported as
``Transient radio bursts from rotating neutron stars'' 
by McLaughlin et al.\ (2006), based on 
a search for isolated bursts of radio emission in data recorded for the 
Parkes multibeam pulsar survey between January 1998 and February 2002.
GASS was completed in November, and 
Pluto was reclassified a minor planet at the IAU General Assembly.

\subsection{2007}
Camilo, Ransom, Halpern, Reynolds (2007) studied
``1E 1547.0-5408: A Radio-emitting Magnetar with a Rotation Period of 2 Seconds,''
using Parkes to discovery the magnetar's period and the ATCA to pin-point the magnetar's position.
S-PASS (S-band Polarisation All-Sky Survey) observing started,
the upgrade of the Parkes Quarters was completed, and Comet McNaught graced the southern skies.

\subsection{2008}
Ghisellini et al.\ (2008)
studied the correlation between the arrival directions of the highest energy cosmic rays 
(detected by the Pierre Auger Observatory) with the position of the galaxies in the 
HIPASS catalogue to consider
``Ultra-high energy cosmic rays, spiral galaxies and magnetars'',
concluding that spiral galaxies are the hosts of the producers of ultra-high energy cosmic rays. 
This year also saw the Lunaska team first use the Parkes telescope in their related search for
radio Cerenkov emission from ultra-high energy neutrinos interacting with the lunar regolith
and producing nanosecond-duration radio pulses peaking in the 1$\sim$2\,GHz range.

The Fermi gamma-ray observatory was launched,
Quentin Bryce became the first female Governor General and the
Beijing Olympics were held.


\subsection{2009}
This was the International Year of Astronomy,
with the Parkes telescope featuring on a commemorative \$1 coin.
Blind searches for periodicities in Fermi gamma-ray data were
resulting in the discovery of gamma-ray pulsars, and
Camilo et al.\ (2009) used archival Parkes data and Green Bank Telescope observations to
report the 
``Radio Detection of LAT PSRs J1741$-$2054 and J2032+4127: No Longer Just Gamma-ray Pulsars.'' 
CSIRO Astronomy and Space Science was formed,
destructive bushfires burned across Victoria, the emergence of a new H1N1 strain
caused a swine flu pandemic, and Michael Jackson died.

\subsection{2010}

It is likely that another Fermi pulsar paper incorporating Parkes data will end up as the
most cited paper from this year, but it is notable that papers from the next generation of
radio astronomers --- 
``A Radio-loud Magnetar in X-ray Quiescence'' (Levin et al.\ 2010), and 
``12.2-GHz methanol masers towards 1.2-mm dust clumps'' (Breen et al.\ 2010) ---
are also having an appreciable scientific impact.
S-PASS observing ended, and the  
Eyjafjallajokull volcano erupted, significantly disrupting air travel.

\subsection{2011}

This year of course saw the Parkes 50th Celebrations, which included Opera at the Dish, 
attended by Governor General Quentin Bryce.
Parkes telescopes were pictured on the Google banner in Australia on October 31st.
The RadioAstron satellite was launched, Japan was rocked by a major earthquake and tsunami and the world population passed 7 billion.

\section{Discussion}

It is clear from this year-by-year review that the Parkes Observatory 
has produced high impact science in a wide range of fields,
both those anticipated when the telescope was first planned ---
H\,{\sc i}, galactic structure, studies of the LMC \& SMC, SNRs, surveys ---
and those unforeseen ---
quasars, masers, planets, molecular lines, radio recombination lines, pulsars, VLBI. 

Bok (1957) wrote 
``One could readily justify the establishment of observatories with
large telescopes in the southern hemisphere because only there can one
study the Magellanic Clouds'' and high impact papers from the years
1966, 1974, 1984, 1988, 1998 and 1999 have confirmed this.
  
Almost a third of the papers highlighted for each year have involved H\,{\sc i} 
observations, and about a quarter have concerned pulsars.
There is a good mix of galactic and extragalactic, spectral line and continuum,
surveys and single objects. There have been notable contributions
by the 18\,m Kennedy antenna, and VLBI observations, with higher frequency
observations ($\nu >$4\,GHz) constituting about a quarter of the papers.
The number of authors on the highlighted paper for each year is plotted in
Figure~3, demonstrating the move to larger research teams over time.

What factors have contributed to the impact of the Parkes Observatory?
The continual upgrading of dish surface, front-end receivers, and
backend processors is clear, highlighting the importance of the
funding brought in by spacecraft tracking contracts, CSIRO support,
and collaborations with the international user community, not to
mention the enabling contributions toward the telescope's construction.
A large majority of high-impact papers have at least one
CSIRO-affiliated co-author, confirming the belief that local knowledge
and experience help to maximise the effective use of facilities. (And
on the other hand, the fact there are high-impact papers unaffiliated
with CSIRO makes it clear that the Observatory is not a ``closed
shop'' and that documentation and user support give all observers the
chance to do good science.)  The excellence of support staff, both at
the Parkes and at RadioPhysics/ATNF/CASS in Marsfield was referred to
by a number of speakers over the week of the Parkes 50th symposium,
and is undoubtedly an important factor in the productivity of the
Observatory.

As exemplified by the 64\,m telescope's appearance on stamps,
coins, banknotes, television advertisements, and cinema, the Dish is
an Australian icon.

The character in the Molnar cartoon claimed the telescope could look
back millions of years.  We now know it can in fact look back billions
of years, and the highlights from its first fifty years ensure it can
look forward to many more!

\acknowledgments

This research has made extensive use of NASA's Astrophysics Data System.

{\it Facilities:} \facility{Parkes Radio Telescope}.

\end{document}